# Predicting Clinical Outcomes in COVID-19 using Radiomics and Deep Learning on Chest Radiographs: A Multi-Institutional Study

**Joseph Bae[1]\*, Saarthak Kapse[1]\*, Gagandeep Singh[2], Rishabh Gattu[2], Syed Ali[2], Neal Shah[3], Colin Marshall[3], Jonathan Pierce[3], Tej Phatak[2], Amit Gupta[3], Jeremy Green[2], Nikhil Madan[4], Prateek Prasanna[1+]**

1    Department of Biomedical Informatics, Stony Brook University
2    Department of Radiology, Newark Beth Israel Medical Center
3    Department of Radiology, University Hospitals Cleveland Medical Center
4    Division of Pulmonary Critical Care, Department of Internal Medicine, Newark Beth Israel Medical Center
\*    Equal contribution
+    Correspondence: Prateek.Prasanna@stonybrook.edu

**Abstract:** We predict mechanical ventilation requirement and mortality using computational modeling of chest radiographs (CXRs) for coronavirus disease 2019 (COVID-19) patients. This two-center, retrospective study analyzed 530 deidentified CXRs from 515 COVID-19 patients treated at Stony Brook University Hospital and Newark Beth Israel Medical Center between March and August 2020. DL and machine learning classifiers to predict mechanical ventilation requirement and mortality were trained and evaluated using patient CXRs. A novel radiomic embedding framework was also explored for outcome prediction. All results are compared against radiologist grading of CXRs (zone-wise expert severity scores). Radiomic and DL classification models had mAUCs of 0.78±0.02 and 0.81±0.04, compared with expert scores mAUCs of 0.75±0.02 and 0.79±0.05 for mechanical ventilation requirement and mortality prediction, respectively. Combined classifiers using both radiomics and expert severity scores resulted in mAUCs of 0.79±0.04 and 0.83±0.04 for each prediction task, demonstrating improvement over either artificial intelligence or radiologist interpretation alone. Our results also suggest instances where inclusion of radiomic features in DL improves model predictions, something that might be explored in other pathologies. The models proposed in this study and the prognostic information they provide might aid physician decision making and resource allocation during the COVID-19 pandemic.

## 1.    Introduction

Coronavirus disease 2019 (COVID-19), an illness caused by novel severe acute respiratory syndrome coronavirus 2 (SARS-CoV-2), has spread rapidly across the world with over 171 million cases internationally and over 33 million cases in the United States as of May 29, 2021 [1]. Advanced cases of the disease can progress to acute respiratory distress syndrome requiring mechanical ventilation [2–7]. So far, over 3.5 million people have died internationally [1]. The ability to identify patients that might progress to critical illness from initial clinical presentation can better guide clinical management strategies and improve patient outcomes [2–4]. Several studies have demonstrated that radiologic imaging may be useful in this regard [2,7,8].

In the United States chest radiographs (CXRs) are the primary imaging modality for the monitoring of COVID-19 and the American College of Radiology has recommended that computed chest tomography (CT) be reserved only for selected patients with limited specific clinical indications including severe disease [2,5,9]. However, CXRs have lower resolution than CT images and provide 2-Dimensional (2D) rather than 3D representations of the lungs. These features make CXRs more difficult to interpret than CTs. Early reports suggested that radiologist diagnosis of COVID-19 from CXR had a sensitivity of 69% compared to a sensitivity of up to 97% on CT [10,11]. Nevertheless, portable radiography is the preferred and often the only available imaging modality in high-volume hospital settings.

Recent studies have qualitatively described the association of ground-glass opacities and lung consolidations with disease severity and progression on CXR and CT [2,5–7,11,12]. Specifically, the presence of opacities in multiple lobes has been shown to predict severe illness, and several CXR scoring systems have been developed to assess disease severity based upon this premise [2,6,7]. Studies have also evaluated various clinical biomarkers and comorbidities as predictors of disease progression, and there is some evidence that imaging data might complement these models [4–6,13–15]. However, current studies to model clinical outcomes in COVID-19 primarily rely on less commonly used CTs or qualitative analysis of CXRs [2,4–6,10,12]. In this multi-site study, we utilize quantitative computational techniques to better evaluate the role of CXR in predicting patient outcomes.

Computational radiology employs machine learning to interpret medical images. Two general approaches include deep learning (DL) and radiomic analysis [8,16]. DL makes use of neural networks to iteratively learn features from CXRs using convolution operations. Radiomic features are distinct, handcrafted attributes that can be directly related to visual characteristics of an image. While recent studies have used these techniques to study COVID-19, few have applied them to multi-institutional CXR cohorts [8,12,17–21].

In this study, we use computational techniques to identify clinically actionable information from baseline CXRs taken for COVID-19 patients. A baseline CXR refers to any CXR taken on the first day for which CXR data exists for a patient treated for COVID-19 infection. First, we develop a baseline model using radiologist assessment of CXR severity (zone-wise expert scores) to predict mechanical ventilation requirement and mortality in order to determine the efficacy of machine learning approaches. We then employ machine learning classifiers to predict these same patient outcomes using computer-extracted radiomic features from baseline CXR. Our third experiment predicts mechanical ventilation requirement and mortality using DL of patient baseline CXRs. Fourth, we propose a combined DL model using both processed CXRs and corresponding radiomic features to predict clinical outcomes. A novel synergistic approach utilizing radiomic-embedded maps for DL is presented and may provide new interpretations of pre-defined radiomic features. Figure 1 displays a general flowchart of experiments.

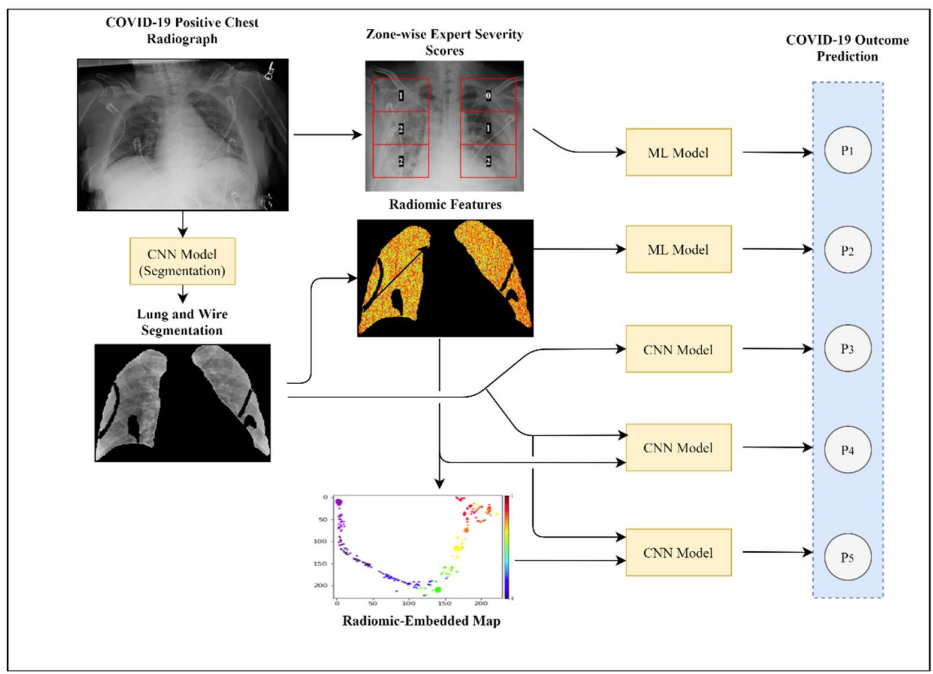

**Figure 1. Study pipeline** Visualized here is the schema for the experiments performed in this study. Experiment 1 demonstrates the use of radiologist expert scoring of CXRs for clinical outcome prediction. In Experiment 2 we extract pre-defined radiomic features from segmented CXRs and input them into machine learning models such as Random Forests. Experiment 3 uses a CNN deep learning model to predict COVID-19 patient outcomes using segmented CXRs as inputs. In Experiment 4 we use extracted radiomic features to generate radiomic-embedded maps which are inputted with segmented CXRs into a CNN deep learning model. Finally, Experiment 5 investigates the value of radiomic feature integration into DL prediction of clinical outcomes using CXRs.

## 2. Materials and Methods

### 2.1. Cohort Description

In this two-center, IRB-approved study, anonymized frontal CXRs were obtained from patients suspected of COVID-19 on presentation at Stony Brook University Hospital (SBUH) and Newark Beth Israel Medical Center (NBIMC) between March and June 2020 (Figure 2). 559 baseline CXRs for 538 patients at SBUH were analyzed. 17 CXRs of pediatric patients or with poor image quality taken from 16 patients were discarded. A total of 174 baseline CXRs from 174 patients were included from NBIMC. Of these, 5 CXRs were discarded due to indistinguishable lung fields. We consider all CXRs taken on the first day for which CXR data exists for a patient as baseline CXRs. Hence, a patient may have multiple baseline CXRs, though these will all be taken on the same day.

In total, 711 CXRs taken from 691 patients (363 males and 328 females) were analyzed in this study. The mean age of patients studied was 56 years old (median=57 years, standard deviation=17.774 years, Table 1). COVID-19 positivity was tested for each patient via reverse transcriptase polymerase chain reaction (RT-PCR). 530 CXRs from 515 patients who tested positive for COVID-19 (Table 2) and 181 CXRs from 176 patients found to not be infected with COVID-19 at SBUH were analyzed. CXRs taken from COVID-19 positive patients were used in outcome prediction experiments whereas those from both COVID-19 positive and negative patients were used to build lung and artifact segmentation models. Of the 530 CXRs from positive patients, 217 baseline CXRs were taken for 205 patients that later required mechanical ventilation. 164 CXRs were from 158 patients who later died from the disease. Representative CXR images are displayed in Figure 3.

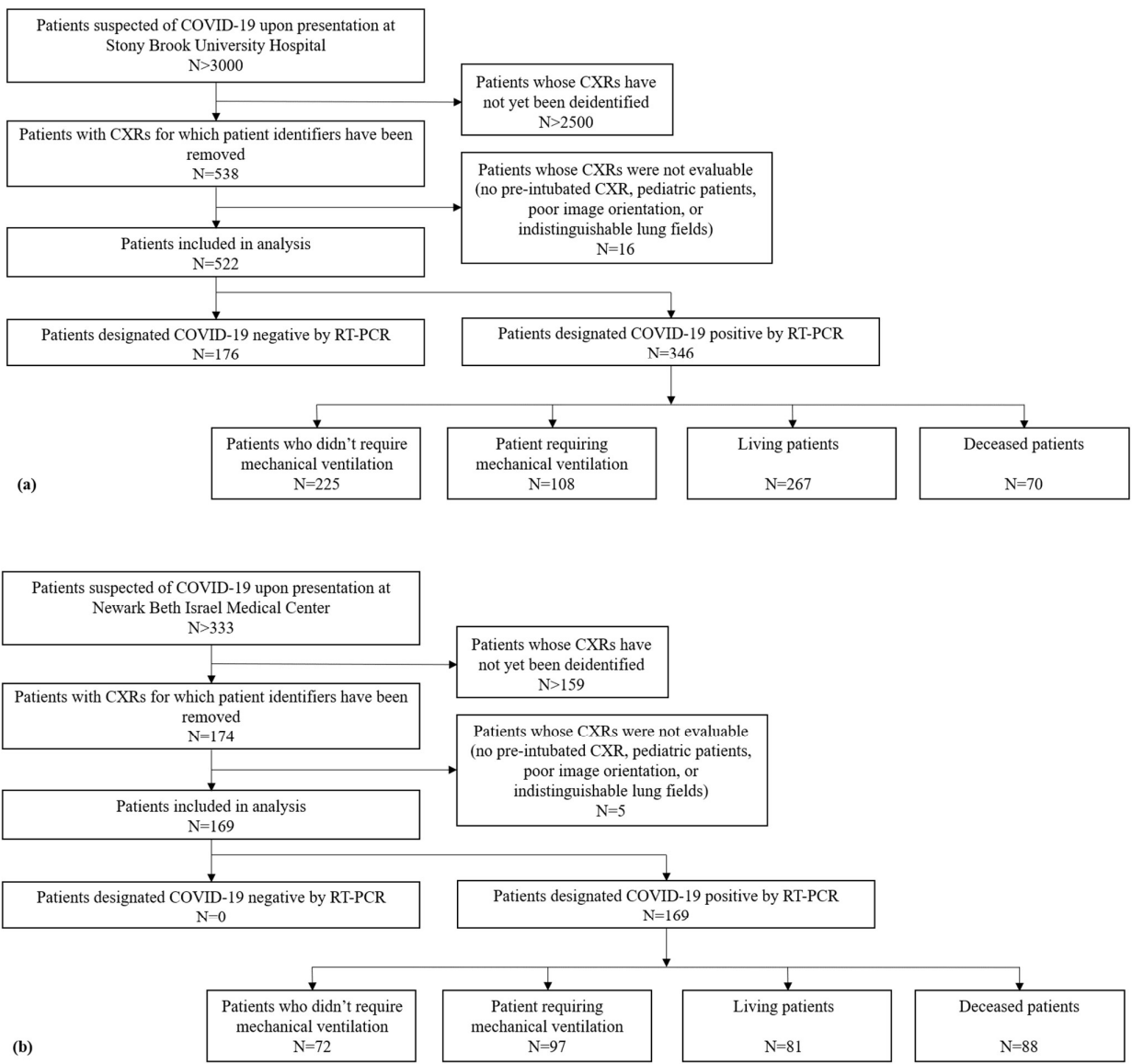

**Figure 2. Summary of patient inclusion and exclusion criteria** (a) displays criteria for SBUH and (b) displays criteria for NBIMC

## Table 1. Total patient demographics table

| | Stony Brook University Hospital patients (N=522) | Newark Beth Israel Medical Center patients (N=169) |
|---|---|---|
| **Sex** | 267 (175 COVID-19+) male 255 (171 COVID-19+) female | 96 male 73 female |
| **Age** | 55±18.630 (p=0.0989*) 57±16.969 (COVID-19+, p=0.1170*) | 59±14.256 (p=0.6821*) |

*p-values for age difference between sexes using a Wilcoxon rank-sum test

**Table 2. COVID-19 positive patient outcome table**

| Age | | Number of COVID-19 positive patients | Number requiring mechanical ventilation | Number deceased |
|---|---|---|---|---|
| **18-19** **(N=1)** | **Male** | 1 | 0 | 0 |
| | **Female** | 0 | 0 | 0 |
| **20-29** **(N=22)** | **Male** | 11 | 1 | 1 |
| | **Female** | 11 | 4 | 2 |
| **30-39** **(N=47)** | **Male** | 29 | 8 | 4 |
| | **Female** | 18 | 4 | 2 |
| **40-49** **(N=75)** | **Male** | 42 | 11 | 5 |
| | **Female** | 33 | 9 | 4 |
| **50-59** **(N=130)** | **Male** | 64 | 24 | 17 |
| | **Female** | 66 | 30 | 13 |
| **60-69** **(N=108)** | **Male** | 60 | 33 | 24 |
| | **Female** | 48 | 27 | 18 |
| **70-79** **(N=79)** | **Male** | 41 | 24 | 20 |
| | **Female** | 38 | 17 | 19 |
| **80+** **(N=53)** | **Male** | 23 | 6 | 11 |
| | **Female** | 30 | 7 | 18 |
| **Total** **(N=515)** | **Male** | 271 | 107 | 82 |
| | **Female** | 244 | 98 | 76 |

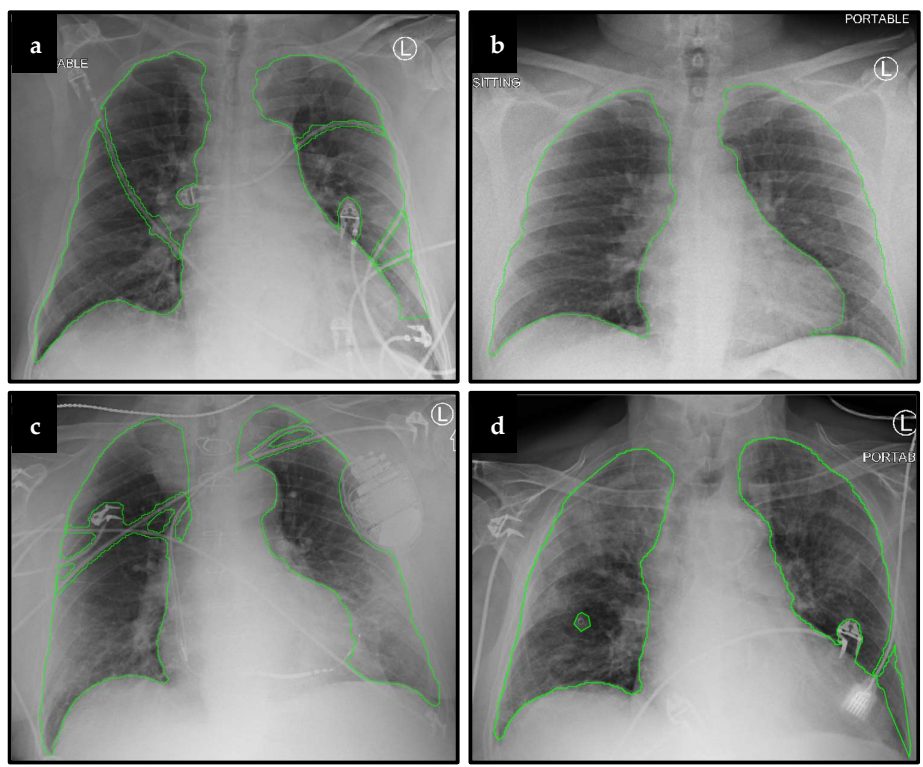

**Figure 3. Representative CXRs studied** Displayed here are baseline CXRs taken from patients that later (a) required mechanical ventilation, (b) did not require ventilation, (c) survived the disease, and (d) did not survive.

## 2.2. Image Preprocessing

### 2.2.1 Lung and artifact segmentation

A segmentation pipeline was developed to avoid analysis of features unrelated to lung fields. In order to segment lungs and artifacts from CXR images, two Residual U-Net DL models were employed [22,23]. Both network architectures were augmented using multiscale image inputs for better intermediate feature representations with deep supervision (Figure 4) [24]. Lung fields and artifacts such as EKG leads, pacemakers, and other non-anatomical objects were first manually segmented for a dataset of 100 CXRs, excluding heart shadows. These segmentations were used to train the two networks, one for lung segmentation and the other for artifact segmentation. A focal Tversky loss function to penalize false positive predictions was employed (alpha=0.3, gamma=1.0) [25]. This was to avoid misidentification of high-intensity objects as lungs and to mitigate misclassification of lungs as unwanted artifacts. The trained models were then used to generate lung and artifact masks for the remaining 611 CXRs. Each of these masks was manually reviewed and errors in segmentation, if any, were corrected.

### 2.2.2 Average Histogram Matching (HM)

It should be noted that CXRs from the two institutions, SBUH and NBIMC, fall within two distinct data domains differing in intensity distribution of pixels. To mitigate image differences, an average histogram matching (HM) was employed (Figure 5). 80 CXR images were chosen randomly from the SBUH dataset to create an average cumulative distribution. Every CXR from both SBUH and NBIMC was then mapped to this average cumulative function using an HM approach, bringing all CXRs into the same intensity range [26].

For both ventilation and mortality classification, models were trained and evaluated in a cross-validation setting. 217 ventilation positive and 300 ventilation negative CXRs were used for ventilation classification, whereas 164 CXRs from deceased patients and 357

CXRs from recovered patients were used for mortality classification. For each iteration of cross-validation evaluation, folds were chosen such that training and testing folds each contained an equal number of positive and negative samples.

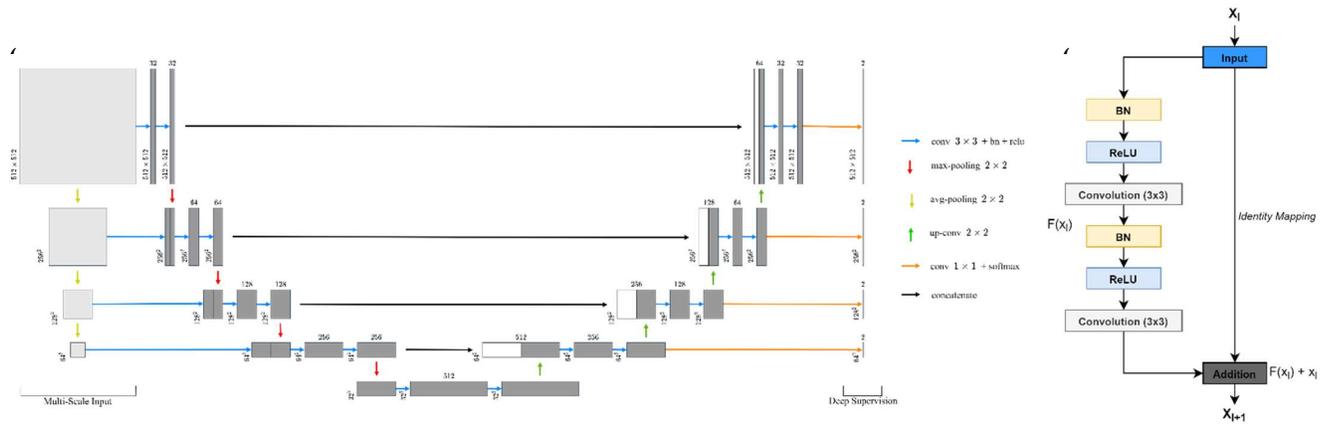

**Figure 4. Network architecture for lung and image artifact segmentation** (a) visualizes our multiscale input Residual U-Net architecture. (b) displays an example residual block.

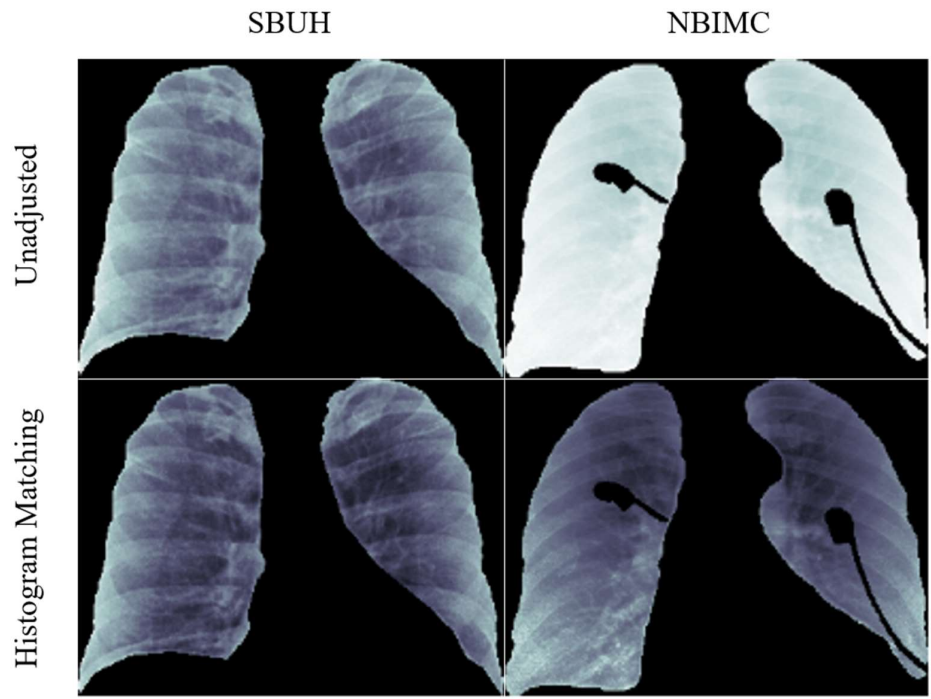

**Figure 5. Results of histogram matching** Displayed are the results of HM preprocessing on CXR images from SBUH and NBIMC

### 2.3. Experiment 1: Outcome classification using radiologist severity scores

In order to develop a clinical baseline model, we adopted a previously described CXR scoring system for COVID-19 patients [7]. Scoring of CXRs was performed by radiology residents (G.S., R.G., S.A., N.S., C.M., and J.P.). Any ambiguous scores were further confirmed by one of two attending radiologists (J.G. and A.G.). For each lung, a severity score of 0, 1, or 2 was assigned to each of three lung zones: lower, middle, and upper (Figure 6), with maximum possible score of 12 for both lungs combined. A score of 0 was assigned to lung zones with no radiographic findings, a score of 1 was assigned to zones with presence of ground glass opacities, and a score of 2 was assigned to zones with consolidative opacities with or without air bronchograms. The formulation of this system and the assignment of different scores to 6 lung zones is in line with other described COVID-19 CXR scoring systems [2,6]. Once these scores were assigned for each CXR, a multiple logistic regression model was developed to predict mechanical ventilation requirement and mortality based upon expert scores. This was evaluated in a cross-validation setting.

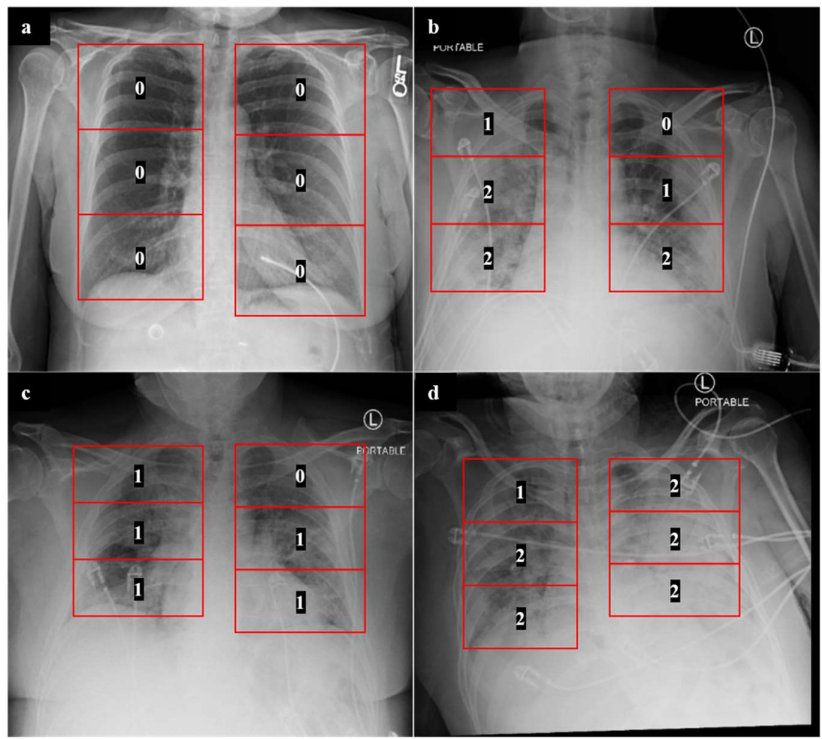

**Figure 6. Zone-wise expert scores for CXRs** Displayed are examples of zone-wise expert scores obtained for COVID-19 patients who (a) did not require mechanical ventilation and recovered, (b) required mechanical ventilation and recovered, and (c)-(d) required mechanical ventilation and are deceased.

### 2.4. Experiment 2: Outcome classification using radiomic features

143 radiomic features from the Haralick, Gabor, Laws energy, histogram of gradients, and grey intensity feature families were computed for each baseline CXR [27–30]. Features were extracted solely from segmented lung fields, excluding artifacts. For each radiomic feature, statistics including measures of median, skewness, standard deviation, and kurtosis were calculated. These statistics and clinical factors including expert scores and patient age/sex were used for classifier construction.

For prediction of future mechanical ventilation requirement and mortality, Random Forest (RF), Linear Discriminant Analysis (LDA), and Quadratic Discriminant Analysis (QDA) classifiers were trained and cross-validated on radiomic features from baseline CXRs [31,32]. For each of 50 iterations in a 5-fold cross-validation setting, feature reduction among radiomic and clinical features was performed on the training set using a Wilcoxon rank sum test, student's t-test, or a maximum relevance minimum redundancy approach [33]. Highly correlated features (Pearson correlation threshold=0.9) were removed to reduce redundancy. Ablation studies were performed to assess the relative performance of radiomic classification with and without HM and with and without clinical features.

### 2.5. Experiment 3: Outcome classification using convolutional neural networks

Convolutional neural networks (CNNs) were employed to predict future mechanical ventilation requirement and patient mortality from baseline CXRs. Additional preprocessing steps for DL included automatic cropping of CXRs to a tight boundary around the lungs, resizing input images to 224 x 224 pixels, and the application of min-max normalization to rescale image intensity values between 0 and 1.

For each classification experiment a ResNet-50 pre-trained on ImageNet was utilized [34]. Data augmentation techniques such as flipping, rotation, and translation were used to reduce overfitting. The fully-connected (FC) layer of each architecture was replaced by a custom layer with an input size of 512 by 1 (no clinical variables included) or 520 by 1 (expert scores and patient age/sex included) and an output size of 2 by 1 to match our desired binary classification scheme. The FC layer was trained without the use of pre-trained weights. Dropout layers with a probability of 0.1 were included after FC layers to improve generalizability of classification. For each model, a binary cross-entropy loss function and an Adam optimizer with a learning rate of 0.00001 were used for network training [35]. The learning rate was decreased by a factor of 0.01 after each $10^{th}$ epoch. Models were trained and evaluated in a cross-validation setting in which new training, validation, and testing splits were chosen for each of five iterations.

Class Activation Maps (CAMs) were also generated using network outputs prior to the global average pooling layer in the ResNet-50 architecture. These CAMs enable a degree of visualization of a network's 'attention' in making predictions thereby providing a soft validation of the prognostically relevant regions as determined by the network.

t-Distributed Stochastic Neighbor Embedding (t-SNE) was used to visualize features extracted using ResNet-50 models for mortality and mechanical ventilation predictions using network outputs prior to the final FC layer [36].

### 2.6. Experiment 4: Outcome classification using convolutional neural networks and radiomic-map embedding

DL of radiomic and imaging features was explored using two different approaches.

#### 2.6.1. Feed-forward concatenation of radiomic features

In this approach, the features used for classifier development in Experiment 2 were first normalized to within a range of 0 to 1 before being concatenated to the output of the upsampling layer of the ResNet-50 architecture used in Experiment 3. The following feed-forward layer was then modified to contain 512+$n$ neurons where $n$ is the number of chosen radiomic features for the desired classification problem. If clinical data including expert scores and patient age/sex were also included, the number of neurons was instead 520+$n$. In this experiment, model weights for the initial image feature extractor layers are used from Experiment 3 whereas the weights for the altered feed-forward layer are randomly initialized. The entire model is then trained. This process was identical for both mechanical ventilation and mortality prediction.

*2.6.2. Radiomic-embedded feature maps*

Radiomic features from Experiment 2 were used to create radiomic-embedded feature maps for each CXR. t-SNE (random state=1) was employed to perform feature reduction and to convert radiomic data to a 2D representation [36]. To assess the predictive capability of a model trained using both radiomic-embedded feature maps and CXR images as inputs, the same general procedure employed in Experiment 3 was used. A key difference was a change in the first input convolution filter of the ResNet-50 architecture to receive a 2-channel CXR and radiomic-embedded map input rather than a 3-channel input. All other network configurations are identical to those described in Experiment 2. Dataset splits of each of these classifiers were identical to those detailed in Experiment 2.

## 3. Results

Results for Experiments 1, 2, 3, and 4 are summarized in Table 3, Table 4, and Table 5 and are reported as mean±95% confidence interval based on 5-fold cross validation results.

**Table 3. Expert scores clinical outcome prediction results**

| Classification Type | Sensitivity | Specificity | AUC |
|---|---|---|---|
| Ventilation Requirement | 0.66±0.07 | 0.64±0.06 | 0.71±0.05 |
| Mortality | 0.76±0.09 | 0.64±0.08 | 0.75±0.06 |

**Table 4. Radiomics clinical outcome prediction results**

| Classification Type | Image Adjustment | Clinical Features | Sensitivity | Specificity | AUC |
|---|---|---|---|---|---|
| Ventilation Requirement | Unadjusted | None | 0.64±0.07 | 0.67±0.07 | 0.72±0.05 |
| | | Expert Scores, patient age and sex | 0.67±0.08 | 0.73±0.07 | 0.77±0.05 |
| | Histogram Matching | None | 0.72±0.07 | 0.72±0.06 | 0.78±0.05 |
| | | Expert Scores, patient age and sex | **0.71±0.06** | **0.71±0.08** | **0.79±0.04** |
| Mortality | Unadjusted | None | 0.72±0.09 | 0.72±0.08 | 0.77±0.05 |
| | | Expert Scores, patient age and sex | **0.79±0.07** | **0.74±0.09** | **0.83±0.04** |
| | Histogram Matching | None | 0.70±0.09 | 0.73±0.09 | 0.78±0.06 |
| | | Expert Scores, patient age and sex | 0.77±0.08 | 0.71±0.09 | 0.80±0.06 |

### Table 5. Deep learning clinical outcome prediction results

| Classification Type | Image Adjustment | Sensitivity | | | | | | Specificity | | | | | | AUC | | | | | |
|---|---|---|---|---|---|---|---|---|---|---|---|---|---|---|---|---|---|---|---|
| | | CXR | CLC | REM | REM CLC | RAD | RAD CLC | CXR | CLC | REM | REM CLC | RAD | RAD CLC | CXR | CLC | REM | REM CLC | RAD | RAD CLC |
| Ventilation Requirement | Unadjusted | 0.55±0.09 | 0.63±0.08 | 0.54±0.08 | 0.58±0.09 | 0.63±0.06 | 0.62±0.07 | 0.72±0.03 | 0.66±0.08 | 0.59±0.05 | 0.65±0.07 | 0.69±0.06 | 0.69±0.06 | 0.70±0.07 | 0.69±0.03 | 0.61±0.03 | 0.64±0.02 | 0.70±0.03 | 0.72±0.02 |
| | Histogram Matching | 0.64±0.09 | 0.61±0.01 | 0.68±0.05 | 0.62±0.08 | 0.66±0.04 | **0.67±0.07** | 0.73±0.07 | 0.76±0.05 | 0.63±0.02 | 0.65±0.06 | 0.75±0.06 | **0.78±0.05** | 0.75±0.02 | 0.77±0.02 | 0.71±0.02 | 0.72±0.02 | 0.77±0.03 | **0.78±0.02** |
| Mortality | Unadjusted | 0.56±0.15 | 0.58±0.17 | 0.61±0.14 | 0.59±0.07 | 0.59±0.12 | | 0.72±0.07 | 0.65±0.06 | 0.58±0.08 | 0.65±0.08 | 0.67±0.03 | 0.71±0.02 | 0.70±0.07 | 0.69±0.06 | 0.68±0.02 | 0.69±0.07 | 0.69±0.07 | 0.71±0.04 |
| | Histogram Matching | 0.59±0.13 | 0.67±0.09 | 0.64±0.07 | **0.77±0.07** | 0.69±0.08 | | 0.74±0.04 | 0.71±0.06 | 0.73±0.07 | **0.60±0.09** | 0.76±0.03 | 0.67±0.03 | 0.74±0.04 | 0.75±0.04 | 0.76±0.04 | **0.77±0.01** | 0.74±0.06 | 0.75±0.02 |

*All columns used preprocessed patient CXRs as inputs to DL networks in addition to:

CLC – clinical features including patient age and sex

REM – radiomic-embedded feature maps

RAD – concatenation of radiomic features to DL feature outputs

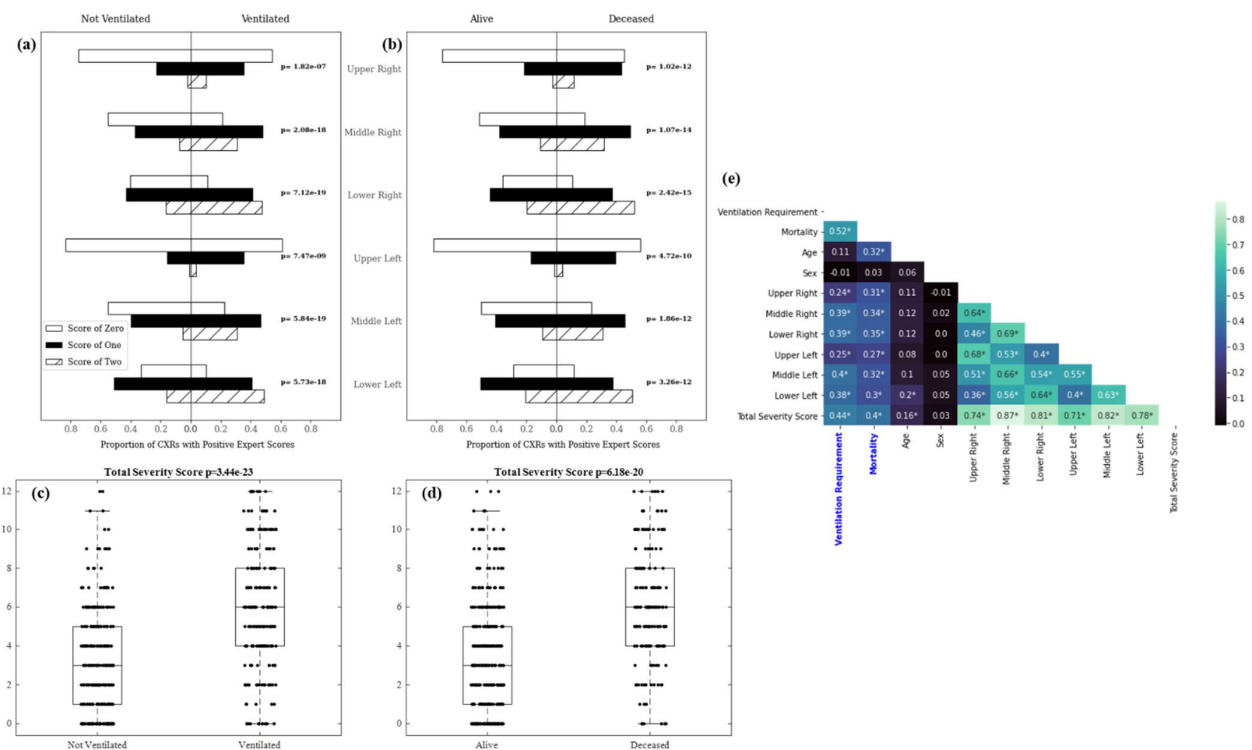

**Figure 7. Zone-wise expert scores distribution** (a) and (b) depict the proportion of patients whose CXRs had lung zone scores of 0, 1, and 2 in each pictured population. (c) and (d) visualize the distribution of total zone-wise scores assigned to CXRs for patients in each population. (e) displays correlations between zone-wise severity scores and clinical outcomes. Asterisks denote statistical significance at the level of p=0.01 as determined by a Pearson's correlation coefficient.

## Table 6. Top 10 features used in radiomic classifiers

| Classification Type | Image Adjustment | Clinical Features | Radiomic Features |
|---|---|---|---|
| Ventilation Requirement | Unadjusted | None | 1. Laws L5E5    2. Gabor XY θ=1.571 λ=1.786    3. Gradient Diagonal    4. Laws E5S5    5. Laws W5E5    6. Laws L5E5    7. Laws W5R5    8. Laws S5E5    9. Haralick Entropy Ws7    10. Haralick Correlation Ws7 |
| | | Expert Scores, patient age and sex | 1. ES Lower Left    2. Age    3. ES Middle Left    4. Sex 5. ES Middle Right    6. Laws W5E5    7. Laws W5R5    8. Laws E5S5    9. Gradient Diagonal    10. ES Lower Right    11. Laws R5E5    12. Laws E5E5    13. Laws E3S3    14. Laws R5W5    15. Laws W5W5    16. Laws S5E5    17. Laws S5W5    18. Laws S5L5    19. Gradient dy |
| | Histogram Matching | None | 1. Gradient Y    2. Laws E5S5    3. Laws L5S5 |
| | | Expert Scores, Patient Age and Sex | 1. Laws E3S3    2. LawsR5R5    3. ES Middle Right    4. ES Lower Right    5. Gabor XY θ=0.785 λ=1.276    6. ES Middle Left    7. ES Lower Left    8. Gabor XY θ=1.963 λ=1.276    9. Grey Standard Deviation 10. Laws L5S5    11. Gabor XY θ=1.178 λ=1.786    12. Haralick Entropy Ws3    13. Gradient Sobel Y    14. Gabor XY θ=1.178 λ=0.765    15. Haralick Information Ws5 |
| Mortality | Unadjusted | None | 1. Haralick Correlation Ws5 |
| | | Expert Scores, Patient Age and Sex | 1. Age    2. Haralick Correlation Ws5    3. ES Middle Right    4. ES Lower Left |
| | Histogram Matching | None | 1. Laws R5E5    2. Gradient Y    3. Laws E3S3    4. Haralick Entropy Ws 5 |
| | | Expert Scores, Patient Age and Sex | 1. Age    2. ES Lower Left    3. ES Middle Right    4. Laws R5E5    5. ES Upper Right    6. ES Lower Right    7. Gradient Y    8. Gradient Sobel YX    9. Laws E3 S3    10. Gradient dx    11. Haralick Entropy |

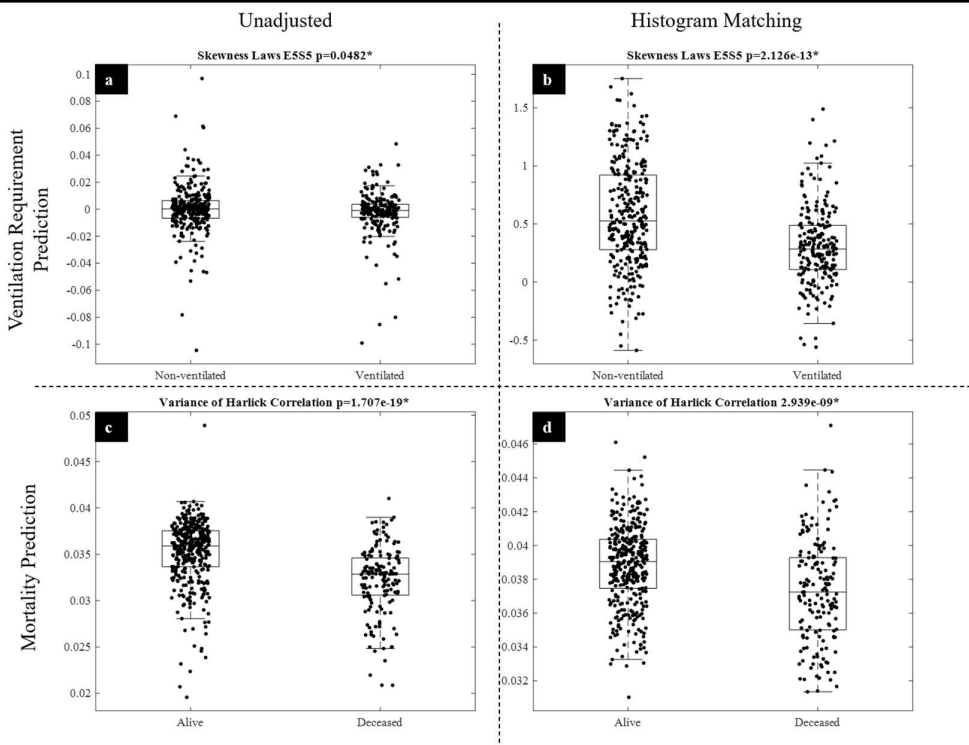

**Figure 8. Radiomic feature distribution** Visualized are the relative effects of HM on the distribution of highly discriminative features for ventilation requirement and mortality prediction.

### 3.1. Experiment 1: Outcome classification using radiologist severity scores

For Experiment 1, expert scores were able to predict mechanical ventilation with a mean cross-validated AUC (mAUC) of 0.75, a specificity of 69%, and a sensitivity of 67%. Expert scores were able to predict mortality with an mAUC of 0.79, a specificity of 76%, and a sensitivity of 69%. Correlations between zone-wise expert scores for each lung region are shown in Figure 7. Total severity score (the sum of scores from all lung regions) correlated most strongly with both future ventilation requirement (0.44) and mortality (0.4), respectively. These correlations were stronger than correlations for individual lung zones with clinical outcomes and for patient age or sex with clinical outcomes.

### 3.2. Experiment 2: Outcome classification using radiomic features

For Experiment 2, a machine learning classifier trained to predict need for mechanical ventilation using radiomic features extracted from non-HM adjusted images yielded an mAUC of 0.72, a specificity of 67%, and a sensitivity of 64%. Using radiomic features from HM adjusted images achieved an mAUC of 0.78, a specificity of 72% and a sensitivity of 72% for mechanical ventilation prediction. A machine learning classifier used to predict mortality in COVID-19 positive patients using radiomic features from non-HM adjusted images had an mAUC of 0.77, a specificity of 72%, and a sensitivity of 72%. Using radiomic features from HM adjusted images resulted in an mAUC of 0.78, a specificity of 73%, and a sensitivity of 70% for mortality prediction. The inclusion of zone-wise expert scores and patient age and sex improved both mechanical ventilation and mortality prediction when combined with radiomic features to yield an mAUC of 0.79, specificity of 71%, and sensitivity of 71% for mechanical ventilation prediction and an mAUC of 0.83, specificity of 74%, and sensitivity of 79% for mortality prediction.

Top features for radiomic outcome classification are listed in Table 6. Among the most discriminating radiomic features identified for predicting mechanical ventilation requirement and mortality were the Laws E5S5 energy and Haralick correlation features, respectively (Figure 8). The Laws E5S5 filter is a composite edge and spot detection filter, whereas the Haralick correlation measures the similarity of a pixel to its neighbors using a grey-level co-occurrence matrix.

### 3.3. Experiment 3: Outcome classification using convolutional neural networks

In Experiment 3, a ResNet-50 model trained solely using non-HM adjusted CXRs to predict future mechanical ventilation requirement had an mAUC of 0.70, a specificity of 72%, and a sensitivity of 55% on cross-validation. Using HM adjusted images as input for DL resulted in improved mechanical ventilation requirement prediction with an mAUC of 0.75, a specificity of 73%, and a sensitivity of 64%. A ResNet-50 model trained using non-HM adjusted CXRs to predict mortality yielded an mAUC of 0.72, a specificity of 72%, and a sensitivity of 56%. Using HM adjusted images for DL training resulted in improved mortality prediction with mAUC of 0.75, a specificity 74%, and a sensitivity of 59%.

### 3.4. Experiment 4: Outcome classification using convolutional neural networks and radiomic-map embedding

For Experiment 4, we found that inclusion of radiomic features improved DL prediction of both mechanical ventilation and mortality. DL models trained using radiomic-embedded feature maps improved prediction of mortality over DL of CXRs alone but did not increase performance when predicting mechanical ventilation requirement. Using feed-forward concatenation of radiomic features to DL features, our model obtained an mAUC of 0.77, a specificity of 75% and a sensitivity of 66% for mechanical ventilation requirement prediction. Using radiomic-embedded features a DL model produced an mAUC of 0.74. a specificity of 76%, and a sensitivity of 59% for mortality prediction. Inclusion of clinical features including expert scores and patient age/sex improved predictions for mechanical ventilation requirement with an mAUC of 0.78, a specificity of 78%, and a sensitivity of 67%. For mortality prediction, inclusion of clinical features improved model predictions to obtain an mAUC of 0.77, a specificity of 60%, and a sensitivity of 77%. Ultimately, inclusion of radiomic features improved DL prediction of clinical outcomes (Table 5).

For DL experiments, representative CAMs are shown in Figure 9. An expert reader (J.G, 15 years of experience) noted that for CXRs from patients that required mechanical ventilation, CAM maximal signal intensity was shown to correlate with areas of dense infiltrates. For selected CXRs for patients who did not require mechanical ventilation, CXRs appeared to demonstrate no focal consolidation or infiltrates. The maximal CAM signal for these CXRs was observed in left middle lung zones, predominantly along the perihilar region. For all CAMs generated, network activations were shown to be most significantly located within lung fields. t-SNE feature reduction for deep features are also visualized in Figure 9. Clustering for features from patients that did and did not require mechanical ventilation is observed.

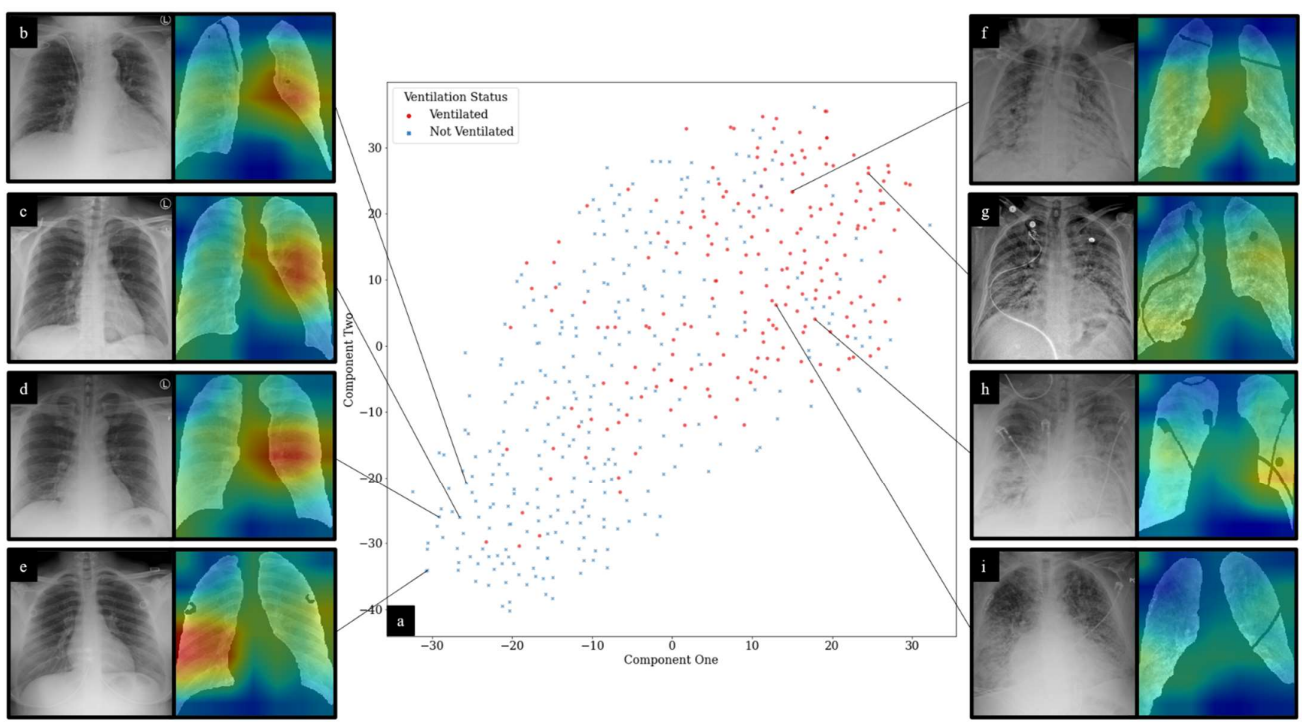

**Figure 9. t-SNE and CAM visualization of DL predictions** (a) displays t-SNE clustering of DL network outputs for ventilation prediction. b)-d) demonstrate no focal consolidation or infiltrates. CAMs show maximal signal intensity in the left middle lung zone predominantly along the perihilar region. e) shows no focal consolidation or infiltrates. CAM shows maximal signal intensity in the right mid to lower lung zone. f) demonstrates diffuse patchy infiltrates bilaterally, predominantly in the mid to lower lung zones. CAM shows highest signal intensities in the right mid to lower lung zones in areas of dense infiltrates. Also noted is slightly increased CAM activity in the left lower lobe around the areas of dense infiltrates. g) demonstrates diffuse patchy infiltrates bilaterally. CAM shows highest signal intensities in the right lower and left upper lung zones around areas of slightly dense infiltrates. h) shows diffuse infiltrates bilaterally with relative sparing of the right upper lobe. CAM shows highest signal intensities in the right mid and left mid to lower lung zones in areas of dense infiltrates. i) demonstrates diffuse bilateral reticular opacities with interlobular septal thickening along with superimposed dense infiltrates predominantly in the lower lobes. CAM shows highest signal intensity in the right lower lung zone around areas of dense infiltrates. CXR interpretation performed by J.G. (15 years experience).

## 4. Discussion

As the COVID-19 pandemic continues to unfold, there will be a growing need for CXR analysis. Several non-imaging models have been proposed with high sensitivities for various clinical outcomes using biomarkers such as serum lactate dehydrogenase, lymphocyte counts, and coagulation factors [4,13–15]. These models might be complemented by imaging-based approaches. Quantitative modeling of baseline CXRs for COVID-19 is being further explored for prognostic tasks and will be of particular importance for resource allocation in high-volume hospitalization settings. In this work we have presented models for baseline CXR analysis demonstrating high sensitivities for future mechanical ventilation requirement (71%) and mortality (79%) prediction. These models outperform expert score-based classification that yield sensitivities of 67% and 69% for mechanical ventilation requirement and mortality, respectively.

Previous studies have applied DL to the analysis of COVID-19 CXRs [8,17,18,37]. However, at least one study has reported potential deficiencies in these approaches, including insufficiencies in a commonly used public dataset, a neglect to segment lung fields, and a failure to account for large differences between disparate public datasets [37]. Most significantly, there has been some suggestion that some studies on a large multi-institutional public dataset may have produced models that learn to distinguish between data taken from different institutions rather than distinguishing meaningful differences in underlying pathology [37]. Nevertheless, new evaluation methods and improvements in data quality might improve experiments performed on these public datasets [38]. Previous studies have also not explicitly accounted for foreign objects in lung fields which can obscure pathological findings. Here, we have further presented a method for dataset homogenization between two separate institutions using HM, addressing any potential discrimination between datasets by our models. Furthermore, we have developed a unique CXR pre-processing pipeline to segment lungs and artifacts.

Radiomic features can provide insight into what characteristics of a patient's CXR are significant in making clinical predictions and can be more informative to a physician than exclusively DL approaches. From our results, it can be observed that radiomic features play an interesting role in outcome prediction for COVID-19. A small subset of radiomic features was shown to be effective in predicting outcome for both mechanical ventilation requirement (3 features) and mortality (1 feature). Radiomic feature classification of future mechanical ventilation requirement improved with HM while also reducing the number of features required for accurate outcome prediction (10 vs 3 features). Interestingly, the opposite effect was observed for mortality prediction; the number of features needed for outcome prediction increased following HM (1 vs 4 features). For ventilation prediction, classifier performances improved following HM whereas HM slightly worsened mortality prediction performance. Laws Energy filters appear to be important in making mechanical ventilation requirement predictions, and Figure 8 demonstrates the observed improvement in Laws E5S5 feature discrimination between classes following HM. For mortality prediction, Laws Energy filters are also selected as discriminatory features following HM. However, the performance of these features in predicting mortality is not as strong as the use of Haralick features prior to HM. Notably, the Haralick Correlation feature does not seem to be "improved" by HM and becomes less valuable in class discrimination for mortality prediction (Figure 8). The variable effect of post-processing techniques on different radiomic feature families warrants further exploration in future experiments. Here we show that two different feature families (Haralick and Laws Energy) might have unique roles in predicting different clinical outcomes and might be variably affected by HM.

In this work we also explore the relative value of two methods of radiomic feature inclusion in deep learning: radiomic feature embedding and feed-forward concatenation of radiomic features. Notably, inclusion of radiomic features improved DL predictions for both clinical outcome tasks. For mechanical ventilation requirement prediction, feed-forward radiomic feature concatenation was superior to radiomic feature appending. The opposite was observed for mortality prediction. This again indicates that different machine learning approaches and selective model invocation may be required for different clinical prediction tasks. We also found that HM uniformly improves DL prediction of clinical outcomes.

In this work we have also demonstrated that radiomic and DL analysis of CXRs can achieve competitive or superior results in predicting clinical outcomes when compared with expert scoring of CXR severity. This is of particular significance in high-volume or low resource healthcare settings where expert annotations may be harder to obtain. Moreover, the combination of DL and radiomic approaches with zone-wise expert scoring of CXRs performs even more accurately in the outcome prediction task, indicating that the two might be applied synergistically to further improve predictions. Furthermore, our models have demonstrated validity on a multi-institutional dataset and might provide a more consistent method of CXR evaluation than human scoring.

There are certain limitations in our work. First, we used baseline CXRs that are likely to be nonuniform in the interval between COVID-19 infection and image acquisition. While this is representative of the clinical reality that patients receive baseline CXRs at varying timepoints in their disease course, future studies might build improved time-to-event prediction models using data with a more uniform temporal distribution. Furthermore, we are limited in the number of clinical features studied and our models might benefit from including co-morbidities such as a history of cancer, chronic obstructive pulmonary disease, hypertension, etc. Finally, additional validation is necessary to demonstrate the robustness of classification models in the broader context of COVID-19 treatment in other hospitals and locations.

This work along with several other recent studies have established the value of computational analysis of CXRs in order to study clinical outcomes in COVID-19 [2,20,21,39,40]. In most cases, these studies analyze CXRs taken at a single timepoint, though modelling of sequential CXR data might enable improved analysis of the temporal evolution of COVID-19 as observed on imaging data.

## 5. Conclusions

In summary, we have presented a complete pipeline for computational evaluation of CXR in COVID-19 patients. Both radiomic and DL classification models enable us to predict mechanical ventilation requirement and mortality from baseline CXRs. Each of these approaches outperforms or performs competitively with predictions made using expert severity assessment of CXRs, indicating the potential for increased efficacy and efficiency in modeling COVID-19 outcomes using machine learning approaches. Furthermore, we demonstrate the improvement that a novel radiomic embedding approach has on DL predictions of COVID-19 outcomes. The ability to make early predictions of disease outcomes may aid in triage, clinical decision-making, and efficient hospital resource allocation as the COVID-19 pandemic progresses.

**Author Contributions:** Conceptualization, P.P., G.S., and J.B.; methodology, J.B., S.K., and P.P.; software, S.K. and J.B.; validation, P.P., J.G. and A.G.; formal analysis, J.B., S.K., G.S., R.G., S.A., N.S., C.M., and J.P.; investigation, J.B., S.K., G.S., R.G., S.A., N.S., C.M., and J.P.; resources, P.P., J.G., T.P., A.G., and N.M.; data curation, P.P., J.G., T.P., A.G., N.M., J.B., S.K., G.S., R.G., S.A., N.S., C.M., and J.P.; writing—original draft preparation, J.B. and S.K.; writing—review and editing, P.P., J.G., T.P., A.G., N.M., J.B., S.K., G.S., R.G., S.A., N.S., C.M., and J.P.; visualization, J.B. and S.K.; supervision, P.P., J.G., and A.G.; project administration, P.P., J.G., and A.G.; funding acquisition, P.P. All authors have read and agreed to the published version of the manuscript.

**Funding:** Research reported in this publication was funded by the Office of the Vice President for Research and Institute for Engineering-Driven Medicine Seed Grants, 2019 at Stony Brook University. JB supported by NIGMS T32GM008444.

**Institutional Review Board Statement** The study was conducted according to the guidelines of the Declaration of Helsinki, and approved by the Institutional Review Boards of Stony Brook University Hospital and Newark Beth Israel Medical Center. All data was deidentified prior to analysis.

**Informed Consent Statement:** All patient data was deidentified prior to analysis.

**Data Availability Statement:** A portion of the data reported in this study will be made available through the Cancer Imaging Archive COVID-19 imaging collection.

**Acknowledgments:** Research reported in this publication was enabled by the Renaissance School of Medicine at Stony Brook University's "COVID-19 Data Commons and Analytic Environment", a data quality initiative instituted by the Office of the Dean, and supported by the Department of Biomedical Informatics.

**Conflicts of Interest:** The authors declare no conflict of interest. The funders had no role in the design of the study; in the collection, analyses, or interpretation of data; in the writing of the manuscript, or in the decision to publish the results.